\documentclass[aps,pra,groupedaddress]{revtex4-1}
\usepackage{amsmath}
\usepackage{graphicx}
\usepackage[usenames,dvipsnames]{color}
\pdfoutput=1

\newcommand{\Eq}[1]{Eq. (\ref{#1})}

\begin{document}


\title{Quantum control and long-range quantum correlations\\ in dynamical Casimir arrays}

\author{Roberto Stassi$^1$}
\author{Simone \surname{De Liberato}$^2$}
\author{Luigi Garziano$^3$}
\author{Bernardo Spagnolo$^1$ $^4$}
\author{Salvatore Savasta$^3$}
\affiliation{$^1$Dipartimento di Fisica e Chimica, Group of Interdisciplinary Theoretical Physics, Universit\`{a} di Palermo and CNISM, Viale delle Scienze, I-90128 Palermo, Italy}
\affiliation{$^2$School of Physics and Astronomy, University of Southampton, Southampton, SO17 1BJ, United Kingdom}
\affiliation{$^3$Dipartimento di Fisica e di Scienze della Terra, Universit\`{a} di Messina, Viale F. Stagno d'Alcontres 31, I-98166 Messina, Italy}
\affiliation{$^4$Istituto Nazionale di Fisica Nucleare, Sezione di Catania, via S. Sofia 64, I-90123 Catania, Italy}




\begin{abstract}
The recent observation of the dynamical Casimir effect in a modulated superconducting waveguide, coronating thirty years of world-wide research, empowered the quantum technology community with a powerful tool to create entangled photons on-chip. 
In this work we show how, going beyond the single waveguide paradigm using a scalable array, it is possible to create multipartite nonclassical states, with the possibility to control the long-range quantum correlations of the emitted photons. In particular, our finite-temperature theory shows how maximally entangled $NOON$ states can be engineered in a realistic setup.
The results here presented open the way to new kinds of quantum 
fluids of light, arising from modulated vacuum fluctuations in linear systems.
\end{abstract}


\maketitle



\section{Introduction}
Generation of photons out of a perturbed vacuum  \cite{Moore1970,FullingDavies1976,Lambrecht77,Uhlmann2004,Crocce05, DeLiberato07,DeLiberato09,Faccio11,Carusotto12,Stassi2013,Garziano2013,Garziano2014,Benenti2014_1,Benenti2014}, usually referred to as dynamical Casimir effect (DCE) \cite{Sassaroli1994,Dodonov10,Dalvit2011} is one of the most fascinating predictions of quantum electrodynamics \cite{Kardar99}, closely related to the better known radiation of Hawking around a black hole \cite{Hawking1975}.

The quest for the observation of such an effect has been one of the great scientific endeavors of our times, successfully concluded with the unambiguous observation of Casimir radiation emitted out of a modulated superconducting microwave resonator \cite{Wilson11,Lahteenmaki13}. Such a groundbreaking result not only definitely vindicated thirty years of study of DCE physics, but it opened a new world for quantum technology research, turning DCE from a theoretical quantum phenomenon into an useful tool for quantum device engineering \cite{Nation2012,Felicetti14}. Casimir radiation can be interpreted as a parametric amplification of the vacuum, creating pairs of entangled photons out of 
a perturbed vacuum \cite{Johansson09}. This makes of DCE an ideal source of entangled photons naturally integrated into superconducting waveguides \cite{Houck2007,Auer12,Johansson2013}. 

In this work we aim to make a major step forward in this direction, proving how it is possible to harness the existing technology of superconducting circuits to both investigate many-body photonic processes and to create nonclassical field states with direct applications in quantum communication devices. We propose a scalable lattice architecture of modulated coupled waveguides in which it is possible to have a fine control over the long-range photonic correlations of the Casimir photons, leading to the possibility to emit on-demand {\em NOON} \cite{Lee2002} and other nonclassical field states and to explore quantum many-body phenomena in coupled linear systems.


\section{Results}

\subsection{The DCE in an array of coupled stripline waveguides}

In this paper we investigate the physics of a one dimensional dynamical Casimir array, that is a one dimensional array made of $N$ superconducting open stripline waveguides (SW), each one terminated on one side by a SQUID loop, threaded by external flux $\Phi_{\text{a}}(t)$ \cite{Johansson2010}. Each terminating SQUID is then coupled to the next one through an additional coupling SQUID threaded by an external flux $\Phi_{\text{b}}(t)$ \cite{Peropadre13}. 

\begin{figure}[hbt]
  \includegraphics[height= 60 mm]{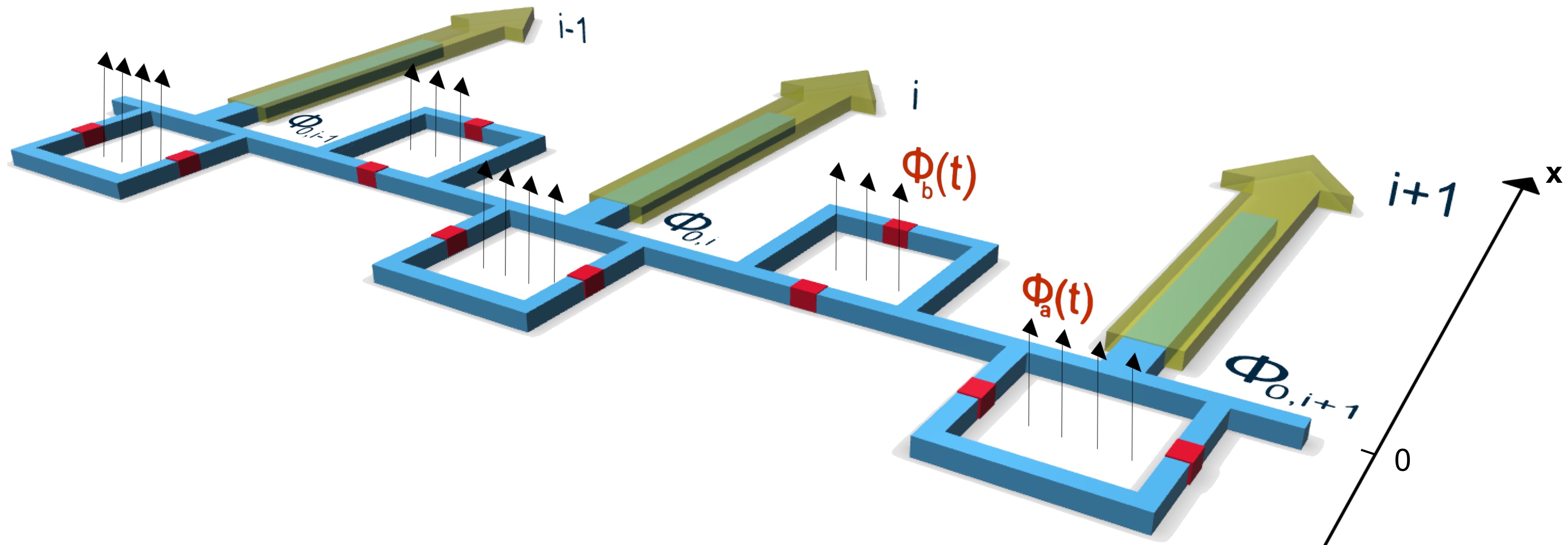}
  \caption{\label{Fig1} Schematic of the array of superconducting waveguides with tunable electric lengths and couplings. The array can have open boundary or can be ring shaped. Each SW is terminated by a SQUID and it is coupled to the nearby SW by a second SQUID. The SQUIDs impose  boundary conditions in the SWs, that can be parametrically tuned by changing the externally applied
magnetic fluxes. We highlighted in red the thin insulating layer of each Josephson junction.  The equivalent circuit diagram for the case of two SWs can be found in Fig.~\ref{Fig2}.}
\end{figure}

This arrangement allows us to both modulate the terminating impedance on each SW, and the coupling between neighboring SWs. We aim to study how the correlations between Casimir photons coming out of the system through the same as well as through different SWs vary as a function of the driving external fluxes.
While a priori each SW could be driven differently, in this paper we will consider both $\Phi_{\text{a}}(t)$ and $\Phi_{\text{b}}(t)$ to be independent from the specific SW. As we will see, this arrangement allows us to have a solid control over the behavior of the quantum correlations whilst limiting to the minimum the number of independent control knobs.
A sketch of the system can be found in Fig. \ref{Fig1}.
It is interesting to notice that arrays of coupled, static quantum circuits (without modulations) have been proposed as promising systems in the field of quantum simulations \cite{Buluta2009,Houck2012,Leib2010}.

\begin{figure}[hbt]
  \includegraphics[height= 80 mm]{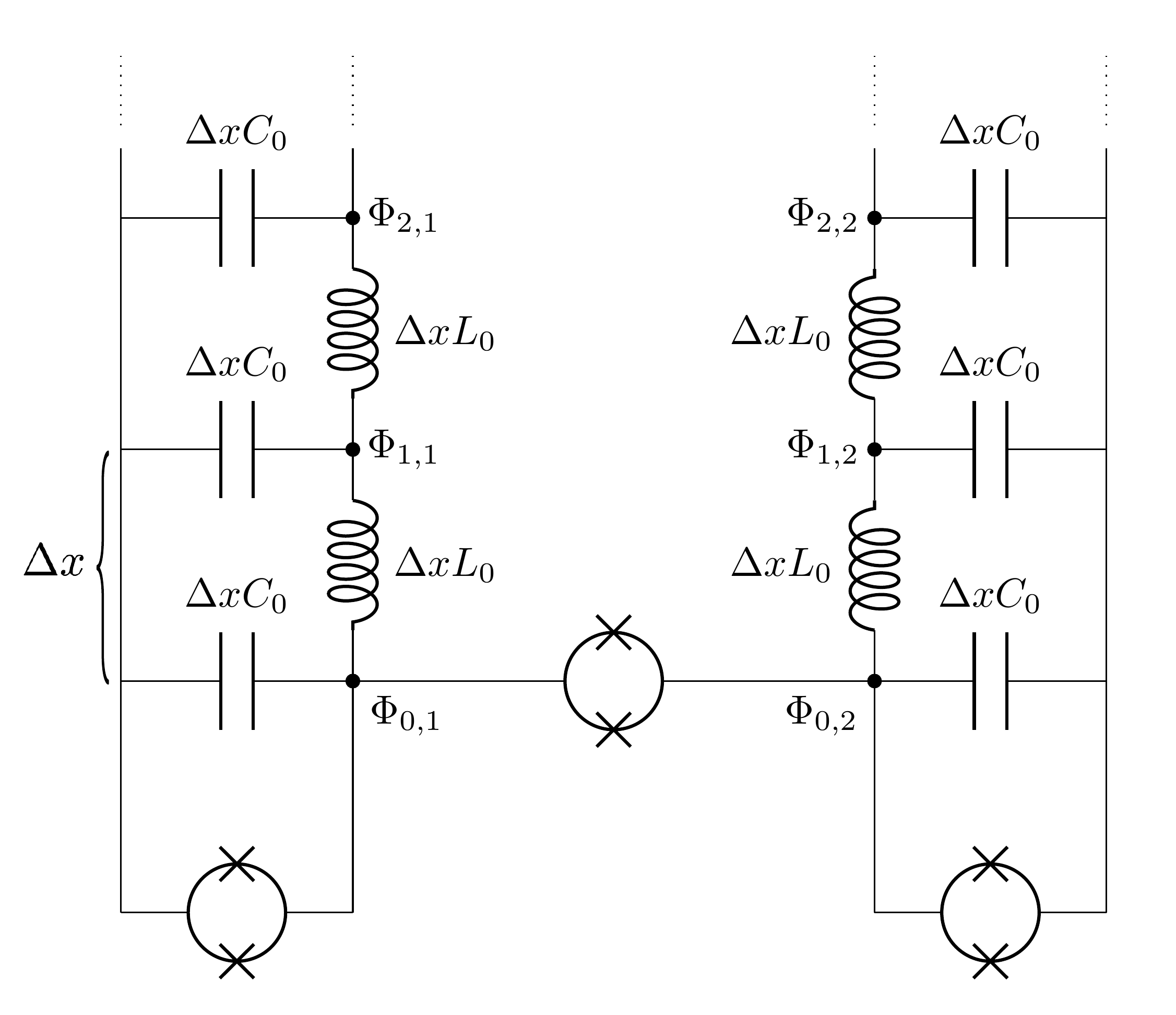}
  \caption{\label{Fig2} Equivalent lumped element circuit diagram for the case of two SWs. The two SWs are characterized by the dynamical fluxes $\Phi_{i,1}$ and $\Phi_{i,2}$ respectively. Each superconducting SW is characterised by its characteristic inductance $L_0$ and capacitance $C_0$ per unit length.}
\end{figure}

Using the standard theory of circuit QED \cite{Yurke84,Devoret1995}, we can identify the dynamical variables of the system as the node fluxes at position $x$ along the $i$-th SW, $\Phi_{x,i}$, and their conjugate momenta $P_{x,i}$ (see Fig.~\ref{Fig2}). The Hamiltonian of the system can thus be written as $H=H_{SW}+H_{SQ}$, where $H_{SW}$ is the Hamiltonian describing the SWs and $H_{SQ}$ the one for the SQUIDs \cite{Johansson2010},
\begin{eqnarray}
\label{H}
 H_{SW}&=&\frac{1}{2}\sum_{i=1}^N\sum_{x =0}^\infty\left [\frac{P^2_{x,i}}{\Delta x C_0}+\frac{\left ( \Phi_{x+1,i}-\Phi_{x,i} \right )^2}{\Delta x L_0}  \right ],\nonumber \\ 
 H_{SQ}&=&\frac{1}{2}\sum_{i=1}^N \frac{P^2_{0,i}}{C_J}+\frac{1}{2}\sum_{i=1}^N E_J\lbrack \Phi_{\text{a}}(t)\rbrack\Phi^2_{0,i}+\frac{1}{2}\sum_{i=1}^N F_J\lbrack\Phi_{\text{b}}(t)\rbrack \left(\Phi_{0,i+1}-\Phi_{0,i} \right)^2.
\end{eqnarray}
In \Eq{H}, $C_0$ and $L_0$ are the characteristic unit length capacitance and inductance of each SW, $E_J\lbrack\Phi_{\text{a}}(t)\rbrack$ and $F_J\lbrack\Phi_{\text{b}}(t)\rbrack$ are the Josephson energies of the terminating and of the coupling SQUIDs, modulated by the respective external fluxes, and $C_{J}$ is the total capacitance in each SQUID \cite{Johansson2010,Likharev}.
Notice that, for sake of simplicity, we neglected additional smaller capacitive terms in the Hamiltonian due to the presence of the coupling SQUIDs and we assumed that the plasma frequency of the SQUIDs far exceeds other characteristic frequencies in the circuit, so that oscillations in the phase across each SQUID have small amplitude.

Our aim is to investigate the correlations among the Casimir photons emitted out of the system through the different SWs. We can thus apply an input-output formalism, effectively tracing out all the internal, unobservable degrees of freedom. In order to do this we extend the procedure described in Ref.~\cite{Johansson2010}, deriving from $H_{SQ}$ a set of boundary conditions over the SWs variables. As detailed in the Appendix A, this procedure allows us to obtain a linear algebraic system linking bosonic ladder operators for photons getting in and out of the $i$-th SW, respectively noted $a^{\rm in}_i$ and $a^{\rm out}_i$. Such a system can be easily solved by expressing the coordinates and momenta for each SWs in terms of collective coordinates and momenta which diagonalize the Hamiltonian in \Eq{H} \cite{Peropadre2013}.
Input and output operators on the $i$-th SW can then be written as linear functions of the corresponding eigenmode operators as 
\begin{eqnarray}
\label{ab}
a_i^{\rm in(out)}=\sum_{n=1}^N c_n^i b_n^{\rm in(out)},
\end{eqnarray}
where $n$ indexes the eigenmodes and the $c_n^i$ are real coefficients.

The SQUID loops provide effective Josephson coupling energies $E_J(t) = E_J[\Phi_a(t)]$ and  $F_J(t) = F_J[\Phi_b(t)]$, tuned by the
threading magnetic fluxes $\Phi_a(t)$ and $\Phi_b(t)$, with a harmonic time dependence. The resulting time-dependent Josephson energies can thus be expressed as
\begin{eqnarray}
E_J(t) = & A_0 \sin (\phi)+\delta A_0 \sin (\theta)\cos\left ( \omega_d t \right )
,\\
F_J(t) = &  A_0 \cos (\phi)+\delta A_0 \cos (\theta)\cos\left ( \omega_d t \right ),\nonumber
\end{eqnarray}
where, in the perturbative regime we consider,  $\delta A_0\ll A_0$. The angles $\phi$ and $\theta$ describe the ratios between the static and variable parts respectively of the terminating and coupling SQUIDs.
The resulting output fields are pairwise correlated, each one with the mode with angular frequency symmetrical around half the driving frequency $\omega_\pm = \omega_d/2 \pm \delta \omega$. 
In the degenerate case $\delta \omega =0$ ($\omega_+ = \omega_- =  \omega_d/2$), the  system can finally be diagonalized obtaining input-output relations for the normal modes of the system
\begin{equation}\label{longeq}
b^{\rm out}_{n}=-b^{\rm in}_{n} - i\frac{\omega_d}{2v }\delta L_n \, b^{\rm in\, \dag}_{n},
\end{equation}
where $v$ is the phase velocity and $\delta L_n$ is a linear function of $\delta A_0$ and depends on $\theta$ and $\phi$  (see Eq.\ (\ref{dL}) and Appendix A). From \Eq{longeq} we clearly see that, as expected, the time-modulation is responsible for the mixing of creation and annihilation operators and thus for the emission of Casimir photons. Notice that the degenerate case, on which we will concentrate in the following, is the most robust against thermal noise, since lower energy modes are exponentially more affected by thermal noise (See Fig.\ 6).
From \Eq{ab} and \Eq{longeq} we can finally find the intensity emitted out of the $i$-th SW as
\begin{eqnarray}
N_i &=&\langle a^{\rm out\, \dag}_{i} a^{\rm out}_{i} \rangle= \left(\frac{\omega_d}{2v} \right)^2\sum_{n=1}^N |c^i_n |^2\, (\delta L_n)^2 \, .
\end{eqnarray} 


Once we have calculated the input-output relations linking input and output channels over the different SWs, we are in measure to calculate correlation functions to extract information on the nature of the Casimir photons emitted in the circuit.
Notice that, while the physically measured quantities are currents and voltages out of each SW, here we will present correlation functions involving directly the single mode photon operators calculated from  \Eq{ab} and \Eq{longeq}. This allows us to deal with dimensionless quantities and standard quantum optics normal ordered correlation functions. Their direct link with correlation functions involving voltage operators is detailed in Appendix E.  Although intensity detectors (measuring normal order correlation functions) in the microwave frequency range are under development \cite{Romero2009,Peropadre11}, it has been shown that these normal order correlation functions can also be inferred by currently used linear detectors \cite{Bozyigit:2011aa}.
The second order correlation function involving fields from SWs $i$ and $j$,  $G_{i,j}^{(2)}= 
\langle a_{i}^{{\rm out}\, \dag}\, a_{j}^{{\rm out}\, \dag}\, a_{j}^{{\rm out}}\, a_{i}^{{\rm out}}\rangle$, for  a vacuum input, can thus be written as
\begin{equation}
\label{G2}
G_{i,j}^{(2)} = \left(\frac{\omega_d}{2v}\right)^2 \sum_{n,m}c^i_n c^i_m c^j_n c^j_m\,  \delta L_n\,  \delta L_m\, .
\end{equation}
The aim of the rest of this paper will be to investigate the very rich physics of \Eq{G2}, and to show how it is possible to exploit it to control long-range photonic correlations and to emit nonclassical multiphoton states.

\subsection{Two coupled stripline waveguides}
We start our analysis from the simplest and analytically solvable case of the DCE array represented in Fig.~\ref{Fig2}, composed of only two coupled SWs, that we will name SW1 and SW2 respectively. 
In this case the transformation coefficients in \Eq{ab} are simply $c^1_1 = c_2^2 = c^1_2 = 1/\sqrt{2}$, $c^2_1 = - 1 /\sqrt{2}$ \cite{Peropadre2013}. We obtain for the intensities,
\begin{eqnarray}
  N_1&=&N_2=\left(\frac{\omega_d}{2v}\right)^2  \frac{ \delta L_1^2+\delta L_2^2}{2}\, .
 \end{eqnarray}

As explained in Appendix B, correlation effects in presence of pairs emission can be better understood considering the intra-SW and iter-SW second order normalized correlation functions defined as $g_{i,j}^{(2)} = G_{i,j}^{(2)} / \sqrt{N_i N_j}$, such that $0\le g_{i,j}^{(2)}\le1$. This normalisation not only allows one to avoid artefacts due to small signals but it also makes the second order correlation function independent from the intensity of the amplitudes $A_0$ and $\delta A_0$, but only dependent on $\phi$ and $\theta$. 
We obtain for the intra-SW and inter-SW second order correlation functions,
\begin{figure}[!ht]
  \includegraphics[height= 100 mm]{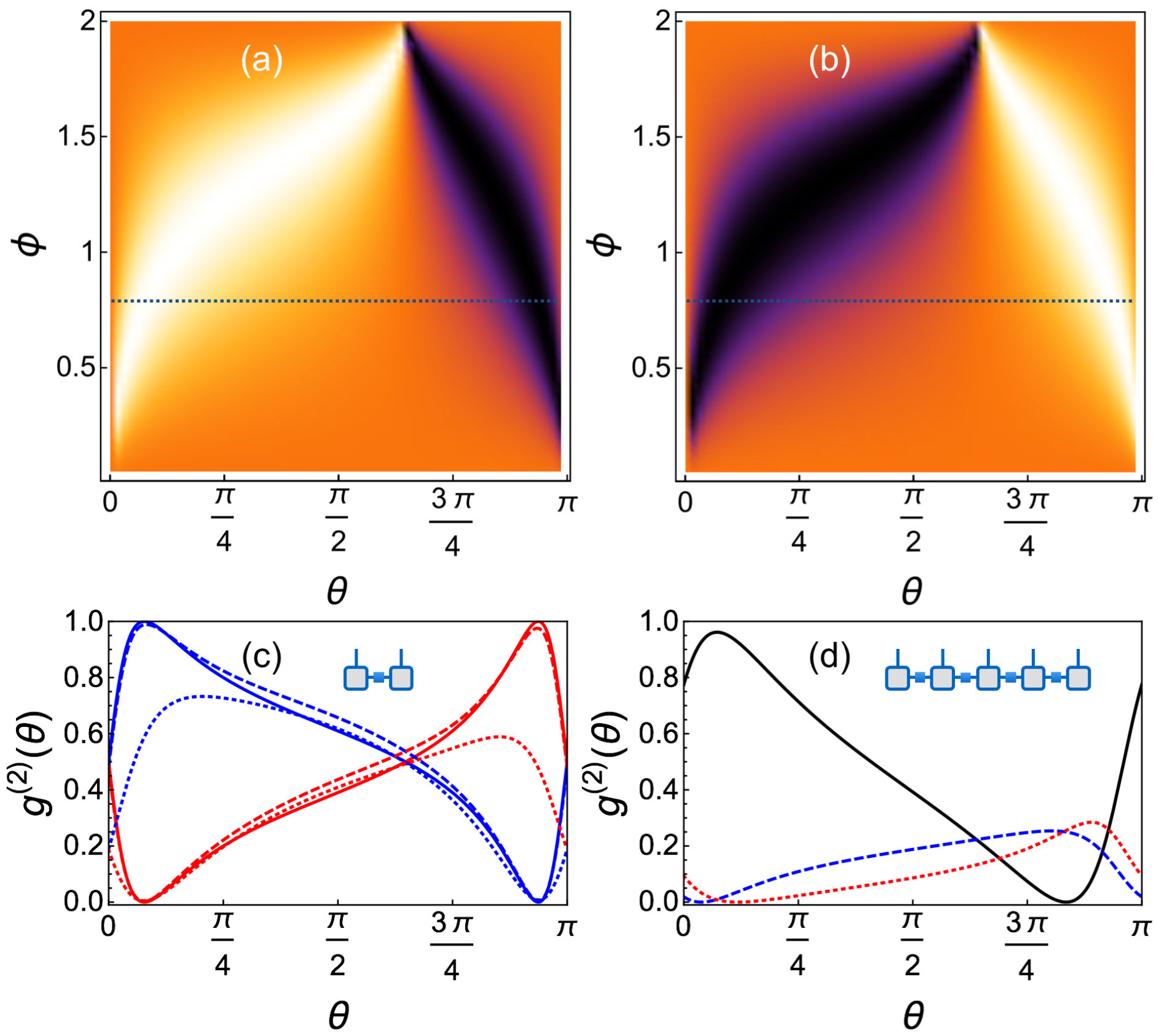}
  \caption{(a) Normalized intra-cavity second order correlation function $g^{(2)}_{1,1}$  for a system of two coupled SWs as a function of  $\theta$ and  $\phi$ for a system of two SWs. (b) Normalized inter-cavity second order correlation function $g^{(2)}_{1,2}$ as a function of $\theta$ and $\phi$ for a system of two SWs. (c) Normalised correlation functions $g^{(2)}_{1,1}$ (red continuous line) and $g^{(2)}_{1,2}$ (blue continuous line) as a function of $\theta $ for  fixed $\phi= \pi/4$. These 1D plots correspond to the line-cuts indicated by dotted lines in Figs.\ 3a and b. The dashed and dotted lines represent instead correlation functions calculated at $T=25$~mK and $T=40$~mK respectively. (d) Normalized second order correlation functions for a system of five coupled SWs with open boundaries: $g^{(2)}_{3,3}$ (black continuous line), $g^{(2)}_{3,1}$ (blue dashed line), and $g^{(2)}_{3,1}$ (red dotted line). Parameter used in the calculations are $Z_0=55$ $\Omega$, $v=1.2\times 10^8\, \text{m/s}$, $\omega_d=2\pi\times 10.3\times 10^9$ and, where not otherwise stated, $T=0$. Moreover, $A_0$ and $\delta A_0$ were chosen so that the maximum value of $\langle a_i^{\rm{out} \dagger} a_i^{\rm{out}}\rangle$ does not exceed $0.1$, at $T=0$.}
\end{figure}
\begin{eqnarray}\label{G11}
  g_{1,1}^{(2)}&=& \frac{1}{2}\frac{ \left ( \delta L_1+\delta L_2 \right )^2}{\delta L_1^2+\delta L_2^2}, \\\nonumber
  g_{1,2}^{(2)}&=&\frac{1}{2} \frac{ \left ( \delta L_1-\delta L_2 \right )^2}{\delta L_1^2+\delta L_2^2},
\end{eqnarray}
where
\begin{eqnarray}\label{dL}
  \delta L_1&=&\left(\frac{\phi_0}{2\pi} \right)^2\frac{1}{L_0}\frac{\delta A_0}{A_0^2}\frac{\sin\theta}{\sin^2\phi}  \, ,\nonumber\\
  \delta L_2&=&\left(\frac{\phi_0}{2\pi} \right)^2\frac{1}{L_0}\frac{\delta A_0}{A_0^2}\frac{\sin\theta+2\cos\theta}{\left(\sin\phi+2\cos\phi\right)^2}  \, ,
\end{eqnarray}
and $\phi_0$ is the flux quantum. 
When either $\delta L_1$ or $\delta L_2$ are zero, $g_{1,1}^{(2)} =  g_{1,2}^{(2)}$, that is, after one photon is detected from the SW1, there is the same probability to detect at zero delay a second photon from either SW1 or SW2. If instead $\delta E_J$ and $\delta F_J$ are chosen so that $\delta L_2 = -\delta L_1$, the probability to get two photons from the same SW vanishes. This photon blockade effect is usually observed in highly nonlinear systems \cite{Birnbaum05,Carusotto2009,Ridolfo12,Rabl11}, while the arrays here described work in the linear regime. Finally, for $\delta L_2 = \delta L_1$, the two photons come from the same SW and the probability to find one photon in each of the two SWs is zero. 
Below we will show that this situation corresponds to the realization of a {\em NOON} state \cite{Lee2002} $|\psi\rangle =\frac{1}{\sqrt{2}} (|2,0 \rangle + |0,2\rangle)$. The system thus works as a switchable source of photons with the unique property to dynamically control the correlation properties of the emitted photon pairs. 

The density plots of $g_{1,1}^{(2)} = g_{2,2}^{(2)}$ and $g_{1,2}^{(2)}= g_{2,1}^{(2)}$ as a function of $\phi$ and $\theta$ can be found in Figs.\ 3a and b. We can see that, for a given $\phi$, changing  $\theta$ allows to modulate $g_{1,1}^{(2)}$ and $g_{1,2}^{(2)}$ over the full spectrum of possible values (Figs.\ 3c and d). This is a demonstration of the full quantum control capability offered by the DCE arrays.
Note that the two plots in Figs.\ 3a and b are linked by the relation $g_{1,1}^{(2)} + g_{1,2}^{(2)} =1$.
Figure 3c shows the line-cuts at fixed $\phi = \pi/4$ of Figs. 3a and b. The angle $\phi = \pi/4$ corresponds to arrays with a stationary coupling potential equal to the stationary potential of the single unit, hence we find that, in the presence of such a large coupling, situations may exists ($\theta = \arctan{(1/4)}$) where photons are bound to their SW and cannot spread in the nearby ones ($g_{1,2}^{(2)}=0$).

Since real experiments are carried out at finite temperatures of the order $T \sim 20 - 60$~mK \cite{Wilson11}, we reported in Fig.\ 3c the inter and intra SW correlation functions at temperatures $T =25$~mK and $T=40$~mK (details on the calculation can be found in Appendix D). These finite temperature calculations demonstrate that the predicted effects are solid and can thus be observed in real experiments under the same conditions used to demonstrate the DCE.  

The obtained intensity correlation functions exhibit interesting quantum features.   
In a two-mode system quantum correlations may exist which violate classical inequalities. If the two modes are symmetric, as in the present case, according to the Cauchy-Schwarz inequality, two fluctuating classical fields (described by a positive Glauber-Sudarshan P function) satisfy the following inequality:
\begin{equation}\label{CS}
g_{1,2}^{(2)} \leq g_{1,1}^{(2)}\, .
\end{equation} 
For $\phi$ values such that $\arctan(-2) \approx 0.65 \pi  \leq\phi \leq \pi$, this inequality is violated (see Fig.\ 2c) hence the system displays nonclassical correlations.
At angles where $\delta L_1 - \delta L_2 = 0$ e.g., $\phi = \pi/4$ and $\theta= \arctan(-1/5)$, there is the maximum violation: no photon pairs in a single SW ($g_{1,1}^{(2)} = g_{2,2}^{(2)}=0$) and perfect inter-SW pair correlation $g_{1,2}^{(2)} =1$. 

\subsection{Beyond two coupled waveguides}
Figure 3d  describes normalized second order correlation functions obtained for an open DCE array composed by 5 SWs. The panel has been obtained for $\phi = \pi/4$ and displays 
\begin{figure}[hbt]
  \includegraphics[height= 80 mm]{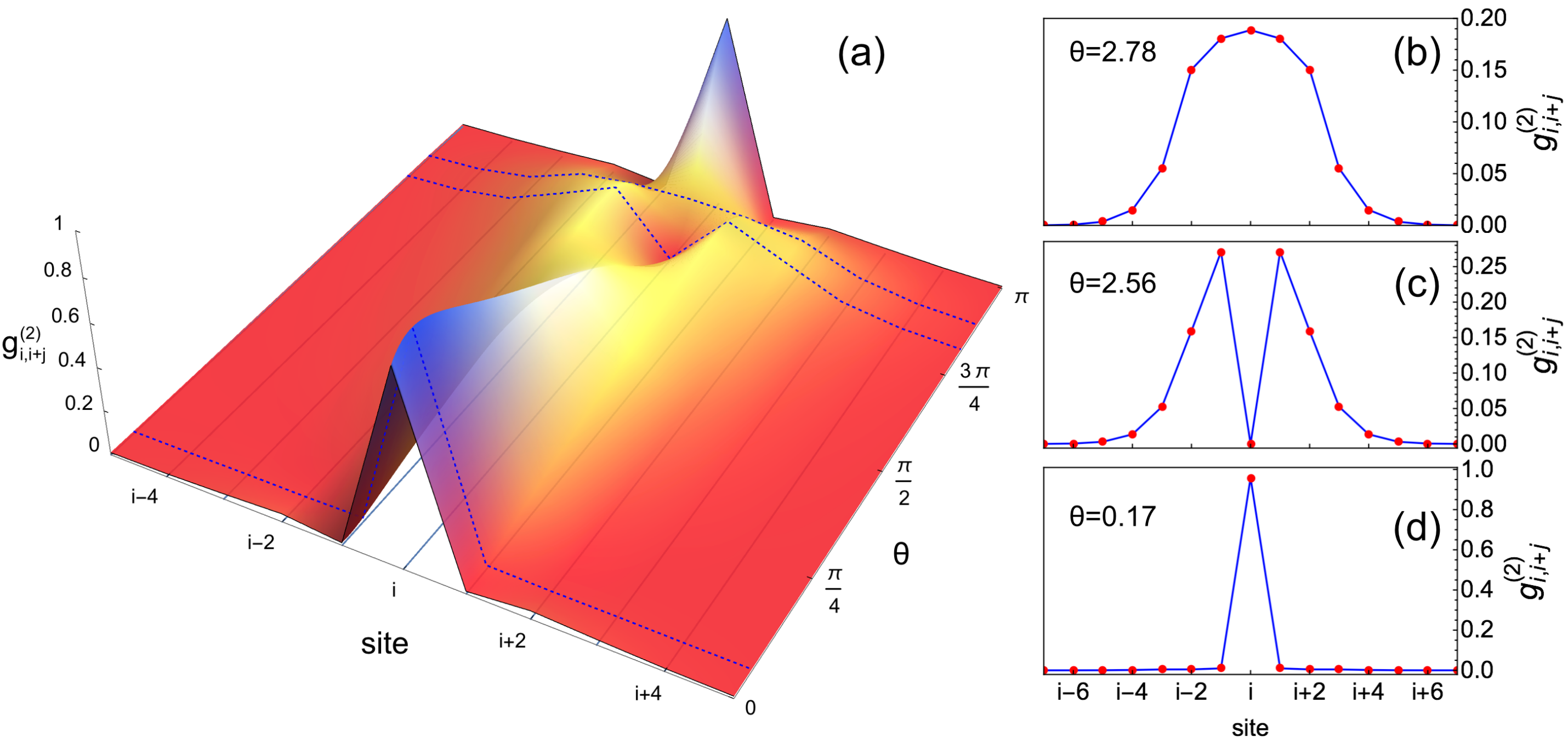}
  \caption{Normalized second order coherence function $g^{(2)}_{i,i+j}$ for ring chain with 31 SWs and $\phi=\pi/4$. (a) 3D representation of $g^{(2)}_{i,i+j}$ in function of site $j=-5,..,+5$ and parameter $\theta$. The blue line-cuts focus on three different scenario for (b) $\theta=2.78$, (c)  $\theta=2.56$, (d) $\theta=0.17$. }
\end{figure}
$g_{3,3}^{(2)}$ (black continuous line), $g_{3,2}^{(2)}$ (blue dashed), and $g_{3,1}^{(2)}$ (red dotted) as a function of $\theta$. We note that the intra-SW correlation remain almost the same as that for the 2 SWs array, and there are regions where both the inter-SW correlations $g_{3,2}^{(2)}$ and $g_{3,1}^{(2)}$ are larger than the intra-SW one $g_{3,3}^{(2)}$. It is surprising to see that there are $\theta$ values where it is more probable to find pairs in more distant SWs rather than in the same SW or in adjacent SWs.
The generated photon pairs show indeed an effective photon-photon interaction that can be tuned from attractive to repulsive simply by acting on the ratio between the modulation amplitudes. However, as discussed above, the possibility to  drastically change the inter- and intra-SWs quantum correlations is actually due to quantum interference effects. 
We have also investigated DCE arrays with a larger number of units. Figure 4 shows results for a ring chain with 31 SWs. In Fig.\ 4\,a  we plot the correlation functions $g_{i,i+j}^{(2)}(\theta)$ for $\phi = \pi/4$ as a function of $\theta$ and $j$. As also shown by the line-cuts obtained for three specific angles $\theta$, the {\em spatial} photon correlations can be controlled to present very different scenarios. Assuming that a photon escapes from the $i$-th SW, the probability of a second photon from any different SW vanishes for $\theta= 0.17$ (Fig.\ 4\,d). This scenario corresponds approximatively (because $g_{i,i}^{(2)}$ is sligthly lower than 1) to the highly entangled multimode {\em NOON} state $|\psi\rangle =\frac{1}{\sqrt{N}}(|2,0,\dots, 0 \rangle + |0, \dots,2,\dots, 0\rangle + 
\dots + |0, \dots, 0, 2\rangle)$.  On the contrary, for $\theta =2.56$ (Fig.\ 4\,c), the probability to find both photons in the same SW is zero. Fig.\ 4\,a shows a still different situation where, for a photon in the $j$ unit, the second one is delocalized around the nearest $8$ units.

\subsection{Entanglement}

The above results provide clear indications that, at specific $\theta$, highly entangled two-photon multimode {\em NOON} states involving all the array units can be realized. In order to confirm this prediction, we have calculated the two-photon array density matrix for two coupled SWs. The matrix elements can be calculated in terms of expectation values involving  SWs output creation and annihilation operators (see Appendix C).
\begin{figure}[hbt]
  \includegraphics[height= 60 mm]{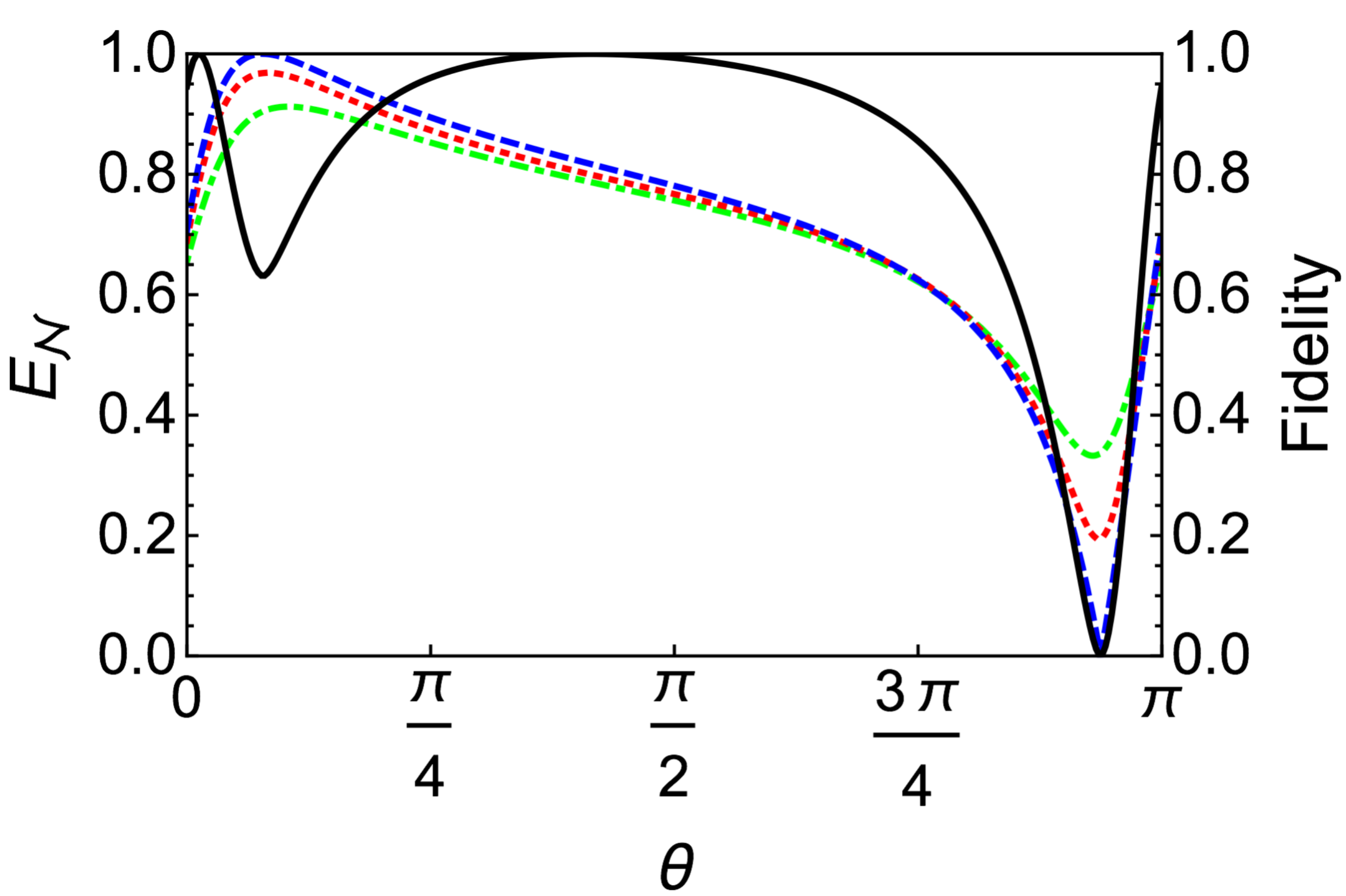}
  \label{EntropyFig}
  \caption{The von Neumann entropy $E_{\cal N}$ (black continuos line) for the reduced density matrix of a system of two coupled waveguides as function of $\theta$ and the fidelity  $\mathcal{F}_{\rm noon}$ of the {\em NOON} state calculated at $T =0$~K (blue dashed line), $T=50$~mK (red dotted line), and $T =60$~mK (green dashed dotted line). The displayed results have been calculated using $\phi = \pi/4$.}	
\end{figure}
Since all the calculations here presented have been obtained perturbatively, in the limit of small modulation amplitudes, in the derivation of the density matrix we considered only those events where at least one photon was generated (post-selection). In this case, each subsystems is constituted by three states corresponding to 0, 1, 2 photons.
In the zero temperature limit the resulting density matrices describe pure states. 
To quantify the entanglement we calculate the von Neumann entropy  $E_\mathcal{N}=-\sum_k p_k\log_3 p_k$  of the reduced density matrix with one of the two SWs traced out, where $p_k$ are the eigenvalues of the reduced density matrix. We considered a 3-basis logarithm so that the maximally entangled state corresponds to $E_\mathcal{N}=1$. While the entropy $E^{\rm tot}_\mathcal{N}$ of the total system is null at $T=0$, indicating that the state is pure, we can see from Fig.\,5 that the entropy of the reduced density matrix is greater than zero for all values of $\theta$ except $\theta =2.94$. It means that $|\psi\rangle$ is an entangled state.  As shown in Fig.\,5, for $\theta=0.038$ and $\theta=1.305$, the entanglement is maximum. This situation corresponds to the generation of the maximally entangled state,
\begin{equation}
  |\psi\rangle= \frac{1}{\sqrt{3}}\left(|1,1\rangle+|2,0\rangle+ |0,2\rangle \right)\, ,
  \label{statepsi}
\end{equation}
where the first (second) entry in the ket indicates the photon number in the first (second) SW. 
When $\theta$ approaches the condition $\delta L_1=-\delta L_2$, so that no photon pairs can be generated in the same SW, the entanglement reduces to zero. However, as certified by the violation of the Cauchy-Schwarz inequality in \Eq{CS}, also in this case the system exhibits highly nonclassical features.

Finally we calculated the fidelity of the {\em NOON} state $\mathcal{F}_{\rm noon}=\sqrt{\langle\rho_{\rm noon} \rangle}$, where $\rho_{\rm noon}= | \psi_{\rm noon}\rangle \langle \psi_{\rm noon}|$ with $ | \psi_{\rm noon}\rangle = \frac{1}{\sqrt{2}}(|2,0 \rangle + |0,2 \rangle)$.
It results (see Fig.\ 5) that for $\theta = \arctan{(-5)}$ corresponding to the condition $\delta L_1=\delta L_2$, the fidelity becomes unitary, demonstrating that a two photon {\em NOON} state is actually obtained. Figure 5 also reports finite temperature calculations at $T =50$~mK and $T= 60$~mK  showing that entanglement results are solid against temperature and the generated  {\em NOON} states can be experimentally observed.

\section{CONCLUSION}

In this paper we studied the physics of an array of SWs in which both the terminating impedances and the inter-SW couplings can be independently modulated in time. We discovered that in this way it is possible to control the long-range quantum correlations of the emitted photons.
In particular, we found it is possible to swipe the second order correlation functions between photons, in the same or in different SWs, over their full range, by tailoring the modulations. We also proved how entangled multipartite {\em NOON} states can be efficiently emitted by the system.

The results here presented open the way to  new kinds of quantum fluids of light \cite{Carusotto2013}, where effective photon-photon interactions arise and can be actively controlled in linear systems. In these systems photons needs not to be externally pumped \cite{Hartmann2006,Houck2012, Leib2010}, but they are generated through quantum vacuum stimulation, so that their correlations can be nonclassical to the highest degree.
Moreover DCE arrays, scalable and versatile by design, naturally allow for control, manipulations via external magnetic fields, and measurements of individual lattice sites. For example, the open waveguides can be replaced  by closed resonators and defects can be easily introduced simply by changing the modulation amplitude of one or more coupling or terminating SQUIDs.
The present work is only a first investigation of the potentialities of DCE arrays. Since the DCE turns vacuum fluctuations into real, observable particles, DCE arrays constitute a test bed for the study of quantum-vacuum correlations in many-body quantum systems.
Future works will have to further explore their possibilities by exploiting their intrinsic design flexibility to produce novel kind of nonclassical many-photon states with long-range correlations when the weak-modulation hypothesis fails. 

\section*{Acknowledgements}

SDL is Royal Society Research Fellow. 
The authors acknowledge support from the Engineering and Physical Sciences Research Council (EPSRC), research grant EP/M003183/1.

\appendix

\section{Output field operators for DCE arrays}
In the continuum limit ($\Delta x \to 0$) \cite{Johansson2010}, the flux coordinate of the $i$-th SW can be expressed as a continuous function of the position: $\Phi_{x, i} \to \Phi_i(x)$.  

In order to derive the output fields, it is useful to find the normal modes of the system transforming coordinates and momenta according to  $\Phi_{i}(x) =\sum_{i}c^{i}_{n}\, \tilde \Phi_{n}(x)$ and $P_{i}(x)=\sum_{i}c_i^n \tilde{P}_{n}(x)$,  such that $H_{S}$ can be expressed in terms of independent oscillators,
\begin{equation}\label{Hsnew}
H_{S}=\frac{1}{2}\sum_{n}\Lambda_n\left(t\right)\tilde{\Phi}_{0,n}^2+\frac{1}{2}\sum_{n}\frac{\tilde{P}_{0,n}^2}{C_J}\, ,
\end{equation}
where the $\Lambda_n(t)$ are linear combination of $E_J(t)$ and $F_J(t)$.
Equation (\ref{Hsnew}) shows that the whole system can be regarded as a collection of independent parametric oscillators.
By writing the Heisenberg equations of motion under the influence of the total Hamiltonian $H$, and taking the continuum limit, the following boundary condition (at $x=0$) can be obtained  \cite{Johansson2010},
\begin{equation}
  C_J \ddot{\tilde \Phi}_n\left(0,t\right) +\Lambda_n\left(t\right) \tilde \Phi_n\left(0,t\right)-\left.\frac{1}{L_0}\frac{\partial \tilde \Phi_n}{\partial x}  \right|_{x=0} =0\, .
  \label{boundcond}
\end{equation}
The SW coordinates $\Phi_i(x)$, as well as the eigenmodes collective coordinates $\tilde \Phi_n(x)$, can be expanded in terms of  input and output creation and annihilation operators \cite{Yurke84}. For example the collective coordinates can be expanded as
\begin{equation}
  \tilde \Phi_n(x,t) =\sqrt{\frac{\hbar Z_0}{4\pi}}\int_{0}^{\infty}\frac{\mathrm{d}\omega }{\sqrt{\omega}}\left [ {\rm b}_{n}^{\rm {in}}\left ( \omega \right )e^{-\imath\left ( -k_{\omega} x +\omega t \right )} +{\rm b}_{n }^{\rm {out}}\left ( \omega \right )e^{-\imath\left ( k_{\omega} x +\omega t \right )}+\rm{h.c.} \right ]\, ,
  \label{qi}
\end{equation}
where $Z_0 = \sqrt{L_0/C_0}$ is the characteristic impedance and $k_{\omega} = \omega/ v$, being $v$ the SW phase velocity. The multimode input and output photon operators obey bosonic commutation relations $[b_{n}^{\rm {in (out)}}(\omega), b_{m}^{\rm {in (out)} \dag}( \omega') ] = \delta_{n,m}\, \delta(\omega -\omega')$.

We now consider applied magnetic fluxes such that $E_J(t) = A_0 \sin (\phi)+\delta A_0 \sin (\theta)\cos\left ( \omega_d t \right )$ and $F_J(t) =  A_0 \cos (\phi)+\delta A_0 \cos (\theta)\cos\left ( \omega_d t \right )$, with weak modulation amplitudes at frequency $\omega_d$.
Inserting Eq.\ (\ref{qi}) into Eq.\ (\ref{boundcond}) and solving perturbatively we can derive the normal mode output fields in terms of the input normal mode operators \cite{Johansson2010}, 
\begin{equation}\label{long}
 {\rm b}^{\rm out}_n(\omega)=-{\rm b}^{\rm in}_n(\omega) + S_n(\omega,\omega_d-\omega){\rm b}^{\rm in\, \dag}_n (\omega_d-\omega)\, ,
\end{equation}
where 
\begin{equation}
\label{temp2}
  S_n\left ( {\omega}',{\omega}'' \right )=-\imath\frac{\delta L_n}{ v }\sqrt{{\omega}'{\omega}''}\Theta \left ( {\omega }' \right)\Theta \left ( {\omega}''  \right ).
\end{equation}
In \Eq{temp2} $\Theta(\omega)$ is the Heaviside function and $\delta L_n=(\phi_0/{2\pi})^2\, \delta \Lambda_n/(L_0{\Lambda^2_{0n}})$, where we have defined
$\Lambda_n=\Lambda_{0n}+\delta \Lambda_n$, with
$\Lambda_{0n}=\lim_{\delta A_0\rightarrow 0}\Lambda_n$.

In order to derive adimensional photon-photon correlation functions it is useful to define single frequency output and input photon operators corresponding to the creation or destruction of photons within a small frequency bandwidth $\Delta$ around a central frequency $\omega_j$ \cite{Johansson09} 
\begin{eqnarray}
{b}_n (\omega_j) = (1/\sqrt{\Delta})\int_{\omega_j -\frac{\Delta}{2}}^{\omega_j +\frac{\Delta}{2}} d\omega\, {\rm b}_n({\omega}),
\end{eqnarray}
obeying standard bosonic commutation relations  $[{b}_n (\omega_j), {b}^\dag_{m} (\omega_{j'})] = \delta_{n,m} \delta_{j,j'}$.
In the main body of the paper we concentrate our attention to the degenerate case with the Casimir photon pairs both at frequency $\omega_d /2$ and thus, for the sake of simplicity, we define
${b}_n (\omega_d/2) \equiv b_n$.

\section{Normalized correlation functions}

The second order correlation function for a generic field $a$ is usually normalized
over the squared population: $\tilde{g}^{(2)}=\frac{G^{(2)}}{N^2}=\frac{\langle \psi \lvert a^{\dagger}a^{\dagger}aa\lvert \psi \rangle}{\langle \psi \lvert a^{\dagger}a\lvert \psi \rangle^2}$, leading to $\tilde{g}^{(2)}=1$ for a perfectly coherent field. This choice of the normalization is nevertheless not well suited to the case of weak processes creating pairs of excitations. The problem can be easily understood considering the state in which parametric process creates a pair of excitations with low probability $\alpha^2\ll 1$: $\lvert \psi \rangle=\lvert 0 \rangle+\alpha \lvert 2 \rangle$, where $\lvert n \rangle$ is the state with $n$ quanta  (such a state is normalized to $O(\alpha^2)$). Direct calculation shows that $\tilde{g}^{(2)}=(2\alpha)^{-1}$, that is, the correlation diverges for a vanishing parametric process. A more physically meaningful way to normalize the second order correlation for these processes, that is the one we employ in the present paper, is $g^{(2)}=\frac{\langle \psi \lvert a^{\dagger}a^{\dagger}aa\lvert \psi \rangle}{\langle \psi \lvert a^{\dagger}a\lvert \psi \rangle}$, that is bounded to unity for states with up to two photons.

\section{Derivation of the density matrix}
In order to calculate the von Neumann entropy and the fidelity of {\em NOON} states to quantify the entanglement, we derived the density matrix $\rho$ for the entire system. 
For a weak driving, a perturbative calculation describing the generation of a single photon pair is adeguate, and the Hilbert space for the two SWs system consists of the tensor product of two qutrits (three-level systems). Each SW can have only 0, 1, 2 photons. The matrix elements $\rho_{nm;n'm'} = \langle |n' m' \rangle \langle n m | \rangle$ (with $n, m, n', m' = 0,1,2$), can be expressed in terms of one and two photon correlation functions already used in the text. For example $\langle |2 0 \rangle \langle 11 | \rangle = (1/\sqrt{2}) \langle (a_1^\dag)^2 a_1 a_2 \rangle$. Since all the calculations here presented have been obtained perturbatively in the limit of small modulation amplitudes, the matrix element $\rho_{00,00}$ is the larger one and we get rid of it by post-selecting only those events where photons are emitted.
\section{Finite temperature}
Finite temperature calculations, displayed in Fig. 3c and 5, can be carried out as the zero-temperature ones, with the only difference that the normal order mean values of the input photon numbers (at frequency $\omega_d /2$) acquire the thermal equilibrium values $\langle b^{in\, \dag}_{n} b^{in}_{n'} \rangle = \delta_{n,n'} N_T$, where $N_T=1/(e^{\hbar\omega_d/2 k_B T}-1)$ is the thermal photon occupation of the input-field mode with frequency $\omega_d/2$, T is the temperature, and $k_B$ the Boltzmann constant.  Higher order input correlation functions can be simply derived employing ordinary Gaussian factorization and using the thermal equilibrium result $\langle  b^{in}_{n} b^{in}_{m}\rangle=0$ \cite{Mandelwolf}.  We obtain for the first order correlation function,
\begin{equation}
  G_{i}^{(1)}=\sum _{n}\left(c^i_n\right)^2\left[N_T+\left(1+N_T \right)\left(\frac{\omega_d}{2v} \right)^2\delta L_n^2\right]\, .
\end{equation}
The second-order correlation function is analogously given by
\begin{equation}
\begin{split}
  &G_{i,j}^{(2)}=
  \sum _{n} \sum _{m} c^i_n c^j_nc^j_mc^i_m(2\, N_T+1)^2\left(\frac{\omega_d}{2v} \right)^2\delta L_n\delta L_m\\
  &+ \sum _{n} \sum _{m}\left(c^i_n c^j_mc^j_nc^i_m + c^i_n c^j_mc^j_mc^i_n \right) \left[N_T+\left(\frac{\omega_d}{2v} \right)^2\delta L_n^2\left ( N_T+1 \right ) \right ]\left[N_T+\left(\frac{\omega_d}{2v} \right)^2\delta L_m^2\left ( N_T+1 \right ) \right ]\, .
\end{split}
\end{equation}

\section{Voltage-based correlation functions}

In microwave circuits, the physically measured fields are the current and voltage along the line. The voltage operator for each SW can be obtained directly from the corresponding flux field through the relation
$V^{\rm out}_i= \partial_t \Phi^{\rm out}_i \left(x,t\right)$. The signal recorded by a photon intensity detector in the $i$-th SW corresponds to the normal order correlation function $G^{(1)}_{i}\left(\tau \right)=\langle V_i^-(\tau)V_i^+(0) \rangle$, where $V_i^+$ and $V_i^-$ are the positive and negative frequency components of the $i$-th SW output voltage $V^{\rm out}_i = V_i^+ + V_i^-$, which can be expressed in terms of annihilation and creation operators respectively.
Although intensity detectors (measuring normal order correlation functions) in the microwave frequency range are under development \cite{Romero2009}, it has been shown that these normal order correlation functions can also be inferred by currently used linear detectors \cite{Bozyigit:2011aa}.
Following this procedure, higher order correlations involving voltages also  from different SWs can be obtained. In the main text, in order to present adimensional normalised correlation functions, we used single mode correlation functions involving photon operators for a small bandwidth around a specific frequency $\omega_d/2$. Here we present voltage-based correlation functions involving fields with a broader spectral range. 
By using the output photon operators derived in the Methods section, it is possible to derive the  first order correlation function \cite{Johansson2010},
\begin{equation}
G^{(1)}_{i}(\tau) =\frac{\hbar Z_0}{4\pi} \int_{0}^{\infty }\int_{0}^{\infty }d{\omega}'d{\omega}''\sqrt{{\omega}'{\omega}''} \langle {\rm a}^{{\rm out} \dagger}_i\left ( {\omega}' \right ) {\rm a}^{\rm out}_i\left ( {\omega}'' \right )  \rangle e^{i\left ( {\omega}'-{\omega}''\right )\tau}\, ,
\end{equation}
where $Z_0 = \sqrt{L_0/C_0}$ is the characteristic impedance. We obtain\begin{equation}
  G^{(1)}_{i}(\tau)=\frac{\hbar Z_0}{4\pi} \sum_{n=1}^{N}\left ( c_n^i  \right )^2\int_{0}^{\omega_d}d\omega\,\omega\left | S_n \left ( \omega,\omega_d-\omega \right )\right |^2\, .
\end{equation}

The photon flux spectral density in the output field, defined by \cite{Johansson2010}
\begin{equation}
  n_i^{\rm out}(\omega)=\int_{0}^{\infty }d{\omega}'\langle {\rm a}^{{\rm out} \dagger}_i(\omega){\rm a}^{{\rm out}}_i(\omega') \rangle
\end{equation}
can thus be expressed as
\begin{equation}
    n_i^{\rm out}(\omega)=\sum_n (c_n^i)^2| S_n \left ( \omega,\omega_d-\omega \right )|^2\, .
\end{equation}
Figure\, 6a displays the photon flux spectral density as a function of the normalized detection frequency ($\omega_d$ is the modulation frequency) calculated for a system of two SWs for $T= 0$ (black solid line), $T = 25$~mK (blue solid line) and in the absence of modulation at $T = 25$~mK (red dashed line). Parameters are provided in the figure caption.

The second order correlation function, defined as $G^{(2)}_{ij}(\tau) =\langle V_i^-(\tau)V_j^-(\tau)V_j^+(0) V_i^+(0) \rangle$ can analogously be written as
\begin{equation}
G_{ij}^{(2)}\left ( \tau \right )=\left(\frac{\hbar Z_0}{4\pi}\right)^2\sum_{n,m}c_n^i c_n^j c_m^j c_m^iI_n(\tau) I_m^\ast (\tau)\, ,
\end{equation}
where
\begin{equation}
  I_n(\tau)=\int_{0}^{\omega_d}\mathrm{d}\omega\sqrt{\omega\left ( \omega_d-\omega \right )}S_n\left ( \omega,\omega_d-\omega \right )e^{i\omega \tau}\, .
\end{equation}
Figure 6 displays $G_{11}^{(2)}(\tau ) / G_{11}^{(2)}(0)$ and $G_{12}^{(2)}(\tau ) / G_{11}^{(2)}(0)$ as a function of the time delay  for a system of two SWs.
The figure shows a decay of these correlations  in the absence of rapid oscillations and without significant modifications of the ratio $G_{11}^{(2)}(\tau ) / G_{11}^{(2)}(\tau)$ when time increases, analogous to what is observed in a single waveguide \cite{Johansson2010}. These results show that the zero delay correlation functions presented in the main text are observable in real experiments where nonzero time-windows are required.
\begin{figure}[hbt]
  \includegraphics[height= 60 mm]{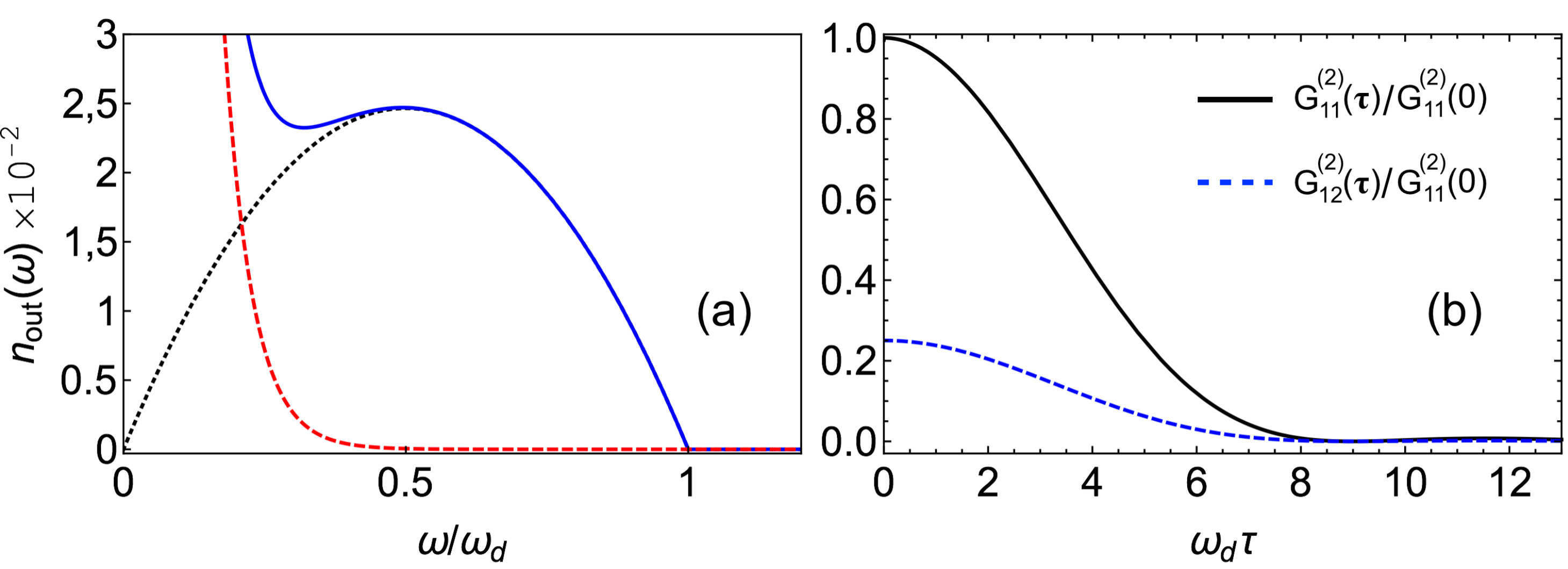}
  \caption{The parameters used in the calculations are $Z_0=55$ $\Omega$, $v=1.2\times 10^8\, \text{m/s}$, $\omega_d=2\pi\times 10.3\times 10^9$. $A_0$ and $\delta A_0$ were chosen so that the maximum value of $\langle a_i^{\rm{out} \dagger} a_i^{\rm{out}}\rangle$ does not exceed $0.1$ at $T=0$. (a) Photon flux spectral density $n_1^{\rm out}(\omega)$ as a function of the normalized detection frequency calculated for a system of two SWs for $T= 0$ (black dotted line), $T = 25$ mK (blue solid line) and in the absence of modulation at $T = 25$ mK (red dashed line). (b) Second order correlation function $G_{11}^{(2)}(\tau ) / G_{11}^{(2)}(0)$ (black line) and $G_{12}^{(2)}(\tau ) / G_{11}^{(2)}(0)$ (blue solid line) as a function of the time delay $\omega_{\rm d}\tau$ for a system of two SWs.}
\end{figure}
\begin{figure}[hbt]
  \includegraphics[height= 60 mm]{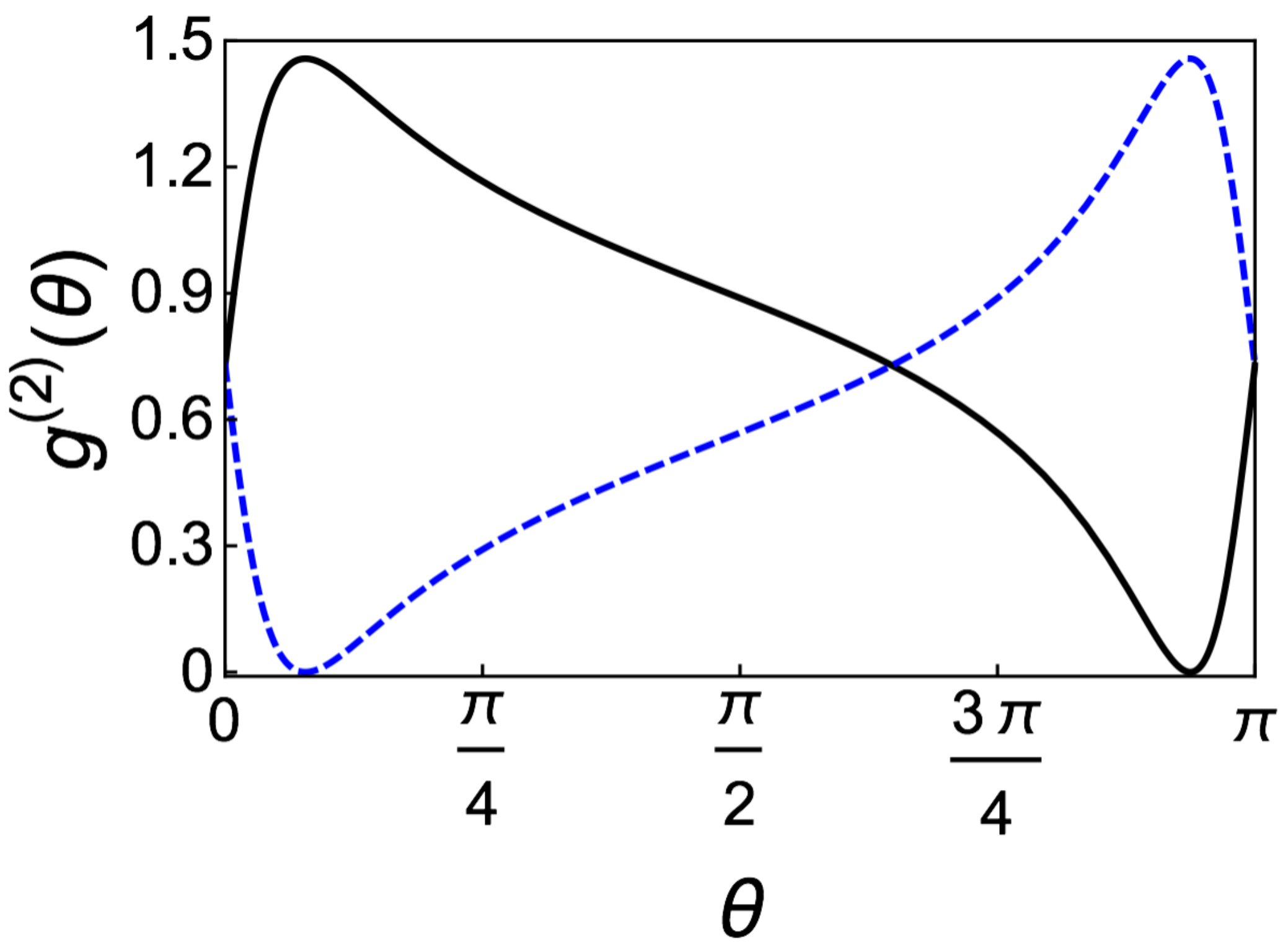}
  \caption{Zero-delay normalized correlation functions $G_{11}^{(2)}(0)/ \sqrt{G^{(1)}_1 G^{(1)}_2}$ and $G_{12}^{(2)}(0)/ \sqrt{G^{(1)}_1 G^{(1)}_2}$ as a function of the parameter $\theta$ are plotted for a system of two SWs at zero temperature.
}
\end{figure}

The zero-delay normalized correlation functions
$G_{11}^{(2)}(0)/ \sqrt{G^{(1)}_1 G^{(1)}_2}$ and $G_{12}^{(2)}(0)/ \sqrt{G^{(1)}_1 G^{(1)}_2}$ as a function of the parameter $\theta$ are reported in Fig.\ 7.
The figure shows that a behavior analogous to the single frequency case showed in the main text (see Fig.\ 3c) can also be observed in broad-band voltage measurement.
The only significant difference is that these normalized correlation functions are no more bounded by 1.

\bibliography{DCEArray}

\end{document}